\documentclass[aps,prl,twocolumn,showpacs,amsmath,amssymb]{revtex4}
\usepackage{graphicx}% Include figure files
\usepackage{dcolumn}% Align table columns on decimal point
\usepackage{epstopdf}
\usepackage{bm}% bold math

\begin{document}
\title{Renormalized spectral function for Co adatom on the Pt(111) surface}
%\title{Exact diagonalization study of Co adatom on the Pt(111) surface}
\author{V.V. Mazurenko$^{1,4}$, S.N. Iskakov$^{1}$, A.N. Rudenko$^{1,2}$, V.I. Anisimov$^{1,3}$ and A.I. Lichtenstein$^{4}$}
\affiliation{$^{1}$Theoretical Physics and Applied Mathematics Department, Ural Federal University, Mira Str.19,  620002
Ekaterinburg, Russia \\
$^{2}$ Institute of Chemical Reaction Engineering, Hamburg University of Technology, Eissendorfer Str. 38, 21073 Hamburg, Germany \\
$^{3}$Institute of Metal Physics, Russian Academy of Sciences, 620219 Ekaterinburg GSP-170, Russia \\
$^{4}$Institute of Theoretical Physics, University of Hamburg, Jungiusstrasse 9, 20355 Hamburg, Germany}
\date{\today}

\begin{abstract}
The strong Coulomb correlations effects in the electronic structure of magnetic Co adatom on the Pt(111) surface have been investigated.  Using a realistic five d-orbital impurity  Anderson model at low temperatures with parameters determined from first-principles calculations we found a striking change of the electronic structure in comparison with the LDA results. The spectral function calculated with full rotationally invariant Coulomb interaction is in good agreement with the quasiparticle region of the STM conductance spectrum.  Using the calculated spin-spin correlation functions we have analyzed the formation of the magnetic moments of the Co impurity orbitals.
\end{abstract}

\pacs{71.27.+a,73.20.At, 75.20.Hr}
\maketitle

Nanosystems consisting of  magnetic atoms deposited on nonmagnetic surface become attractive for the development of novel types of memory and  computation devices. The electronic and magnetic properties of such systems can be probed by means of the inelastic scanning tunneling spectroscopy (STM) \cite{IBM2}.  The STM spectra demonstrate different types of elementary excitations.  For instance, if the magnetic atoms are deposited on insulating surface there are steps structure in the differential conductance $dI/dV$ spectrum \cite{hirjibehedin} that is related with inelastic excitations and can be reproduced in the framework of the spin tunneling Hamiltonian approach \cite{Appelbaum, fernandez}. The latter is based on the fact that well-localized magnetic moments of the adatoms can be described with a simple Heisenberg spin-Hamiltonian \cite{hirjibehedin,rudenko}.

In contrast to the insulating substrate, the experimental STM spectrum  of magnetic adatoms deposited on a metallic surface can show the Kondo-like physics \cite{Kondo,Hewson}. Nevertheless, the recent spin-polarized STM experiments \cite{Meier, Zhou} for Co impurities on Pt(111) surface have demonstrated the existence of narrow peaks near the Fermi level which are not split in the external magnetic field.  Moreover, the width of this peak of around 150 meV is much larger than that one would expect for a Kondo system \cite{Kondo,Otte}. From the theoretical point of view this peak can be related with the local density of state of the magnetic adatom \cite{Kelly}. Indeed, the DFT-based calculations  \cite{shick, weinberger, sabiryanov} have demonstrated a peak near the Fermi level, however, the width of the peak is much larger than the experimental one and is located fare away from zero bias. The main problem with the DFT-calculations is the missing of dynamical quantum fluctuations which can be very important in case of strongly correlated nano-systems \cite{Georges}. They can lead to a considerable renormalization  of the LDA spectrum near the Fermi level \cite{Mazurenko}. In order to capture the local correlations on magnetic adatom properly one should solve the many-body Anderson-type Hamiltonians \cite{Hewson,Gorelov} at experimental temperatures.
\begin{figure}[]
\includegraphics[width=0.38\textwidth,angle=0]{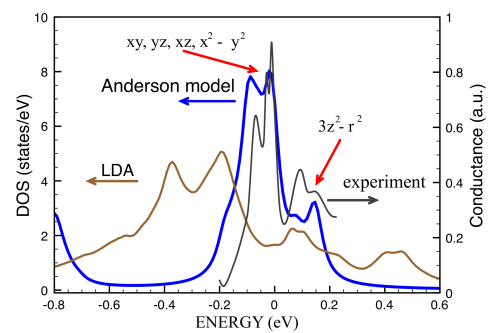}
\caption{(color online). Densities of states obtained from LDA calculation (brown line) and many-body Anderson model calculations (bold blue line) in comparison with of the quasiparticle region of the experimental differential conductance \cite{Meier} (thin gray line).}
\label{structures}
\end{figure}

In this Letter we solve the realistic five d-orbital Anderson impurity model with parameters determined from first-principles calculations to describe the electronic and magnetic properties of Co/Pt(111) system.  Using different parametrization of the Coulomb interaction matrix we have analyzed the microscopic origin of low-energy excitations of the system. Our main result can be in short presented in Fig.1 which gives the comparison between the quasiparticle region of the experimental STM differential conductance \cite{Meier} with the spectral functions obtained from the LDA and temperature-dependent Anderson impurity-model calculations. One can immediately realized that there is a huge renormalization of the quasiparticle density of states in Anderson model in comparison with the LDA results. Moreover a  good agreement between experimental results and Anderson model can be obtained only with the full rotationally-invariant parametrization of Coulomb interaction matrix.  The peaks in the density of states below and above the Fermi level are formed by Co-3d states of different symmetry which can be checked in the experiments. We have also calculated the partial spin-spin correlation functions for different Co orbitals in order to study formation of the magnetic moment.

{\it LDA analysis.}
We performed the LDA calculation with structural relaxation for single Co atom on the Pt(111) surface
by means of the projected augmented-wave (PAW) method \cite{PAW} as implemented in the Vienna {\it ab-initio} simulation package (VASP) \cite{VASP}.
We used an energy cutoff of 300 eV in the
plane-wave basis construction and the energy convergence criteria of
$10^{-7}$ eV. Atomic positions of the considered system were relaxed with
residual forces less than 0.01 eV/$\AA$. To simulate the atomic structure of Co adatom on the Pt(111) surface we have used the supercell approach.
The supercell contains three-layer (3{ }$\times${ }3)
Pt(111) surface, Co atom and vacuum region of 10 $\AA$. Lattice constant
for Pt lattice was chosen to be 3.92 $\AA$ that is the experimental value of lattice
constant for the bulk fcc Pt. The lowermost layer of Pt has been fixed under relaxation.
The obtained vertical distance of Co atom of 1.82 $\AA$ is in agreement with the reported values \cite{sabiryanov,shick}.

Fig.2 gives the LDA partial densities of states.
All the cobalt 3d states can be classified with respect to the hybridization with Pt states.
One can see that the density of states of $3z^2-r^2$ symmetry lies below the Fermi level and demonstrates a localized peak.
It is due to a small overlap and hybridization with states of the nearest Pt atoms.
In turn the states of the in-plane orbitals of $xy$ and $x^2-y^2$ symmetry hybridize with
Pt states and are wider than that of $3z^2-r^2$ orbital. The largest hybridization occurs between $xz$ ($yz$) and 5d Pt states.

\begin{figure}[]
\includegraphics[width=0.29\textwidth,angle=0]{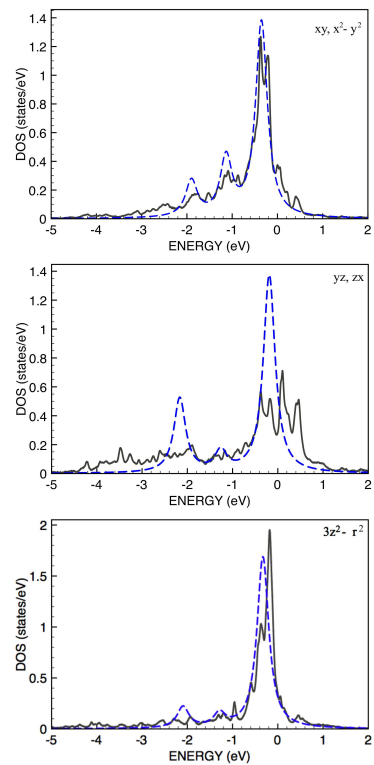}
\caption{(color online). Densities of states obtained from LDA calculations (gray solid line). The blue dashed lines correspond to fitting results obtained by using Eq.(2).}
\label{structures}
\end{figure}

To study the magnetic properties of Co/Pt(111) we have performed spin-polarized LDA calculations.
The obtained value of the local spin magnetic moment $M_S=2.09 \, \mu_B$ is the same
for out-of-plane and in-plane direction of magnetization,
whereas the orbital moment varies from $M_L^z=0.16 \, \mu_B$ to $M_L^x=0.09 \, \mu_B$.
These values are much smaller than the total giant magnetic moment of Co-Pt systems observed experimentally.

{\it Coulomb correlations.}
One can see that in the framework of the LDA the orbital polarization is strongly underestimated.
The same result was obtained in the previous first-principles investigations \cite{weinberger, sabiryanov}. To overcome this problem the different kinds of the orbital polarization corrections were used to fit the experimental data.
Using the static Coulomb correlations within the LSDA+U method the correct  value of the orbital magnetic moment in Co-Pt systems have been reproduced \cite{shick}.

The static mean-field LDA+U and many-body approaches  we use require the on-site Coulomb interaction and intra-atomic exchange interaction as input parameters. To calculate them we have used the constrained LDA procedure \cite{gunn}.
 Our calculations have shown that the values do not differ significantly for relaxed
 (U = 6.6 eV, J = 0.9 eV) and non-relaxed (U = 6.9 eV, J = 0.9 eV) structures. Thus in our many-body calculations we have used the averaged values U= 6.75 eV  and J = 0.9 eV. Using the estimated  U and J we have performed the LDA+U calculations in the framework of the tight-binding linear muffin-tin orbital approach within the atomic sphere approximation (TB-LMTO-ASA) method \cite{Andersen,Anisimov}.
The calculated value of the magnetic moment $M_{S}$ = 2.8 $\mu_{B}$ is larger than those experimentally observed \cite{gambardella1}. It is well known that the account of the many-body correlations can result in a renormalization of the electronic spectrum and
as a consequence the magnetic moments becomes smaller \cite{Mazurenko}.

{\it Many-body calculations.}
The importance of account of the dynamical Coulomb correlations for description of the adatom on a metallic surface was emphasized in Ref.\onlinecite{Gorelov}.
The authors have investigated the electronic structure of a cobalt impurity on the surface of Cu(111) by using a CT-QMC method.
They have found that in contrast to the bulk system there is a large sign problem related with non-diagonal elements of the Coulomb interaction matrix. The sign problem prevented them from calculating the physical properties at low temperatures.

Here we use the finite temperature exact diagonalization Lanczos-method \cite{Capone} in order to solve the Anderson impurity Hamiltonian for realistic multi-orbital systems\cite{Liebsch, Mazurenko}.
We consider such a method to be preferable in comparison with others because one can use the full U-matrix and the simulation can be performed at experimental temperatures \cite{gambardella1}. The main  disadvantage of the exact diagonalization is a discretization of the LDA spectrum required to define the impurity and bath energies as well as  the hopping parameters in the Anderson Hamiltonian.

To study the correlation effects we diagonalized the Anderson impurity Hamiltonian \cite{Anderson}
\begin{eqnarray}
H = \sum_{k \sigma} \epsilon_{k} n_{k \sigma}  + \sum_{m \sigma}   (\epsilon^{\sigma}_{m} - \mu) n_{m \sigma} \nonumber \\
+ \sum_{m k \sigma} ( V_{mk} d^{+}_{m \sigma} c_{k \sigma} + V_{km}c^{+}_{k \sigma} d_{m \sigma})  \nonumber \\
+ \frac{1}{2} \sum_{\substack {m m' m''\\  m''' \sigma \sigma'}}  U_{m m' m'' m'''} d^{+}_{m \sigma} d^{+}_{m' \sigma'} d_{m''' \sigma'} d_{m'' \sigma},
\end{eqnarray}
where $\epsilon_{m}^{\sigma}$ ($\epsilon_{k}$) is the energy of the impurity (surface) states, $d^{+}_{m \sigma}$ ($c^{+}_{k \sigma}$) is the creation operator for impurity (surface) electrons, $V_{mk}$ is a hopping between impurity and surface states, and $U_{mm'm''m'''}$ is the Coulomb matrix element \cite{Gorelov}. Importantly,  in our calculations the full Coulomb interaction vertex Eq.(1) as well as the density-density form $H^{den} =\frac{1}{2} \sum_{mm' \sigma} [U_{mm'mm'} n_{m \sigma} n_{m' -\sigma} + (U_{mm'mm'} - U_{mm'm'm}) n_{m \sigma} n_{m' \sigma}] $   were used.
The chemical potential $\mu$ was fitted to obtain the number of impurity electrons of 7.6.

The energies $\epsilon_{k}$, $\epsilon^{\sigma}_{m}$ and the hopping parameters $V_{mk}$ were found by minimizing
the LDA Green functions  using the following expression \cite{Georges} :
\begin{eqnarray}
G^{LDA}_{m \sigma} (\omega)  = (\omega - \epsilon^{\sigma}_{m} - \sum_{k = 1}^{N_{s}} \frac{|V^{\sigma}_{mk}|^2}{\omega - \epsilon_{k}}).
\end{eqnarray}
In our calculations we used $N_{s}$ = 2 for each impurity orbital, thus the total number of the electronic levels was equal to 15.
The same number of the bath and impurity orbitals was used in Ref. \onlinecite{Liebsch15} to solve a DMFT problem.
Fig.2 gives the comparison between LDA densities of states and fitted spectral functions. One can see that there is good agreement between them.
The obtained values of $\epsilon_m$ agree well with an energy centers of the corresponding orbital.
The smallest hybridization hoppings were obtained for $3z^2-r^2$ orbital, since this orbital is the most localized one.

The impurity Green's functions and spin susceptibilities are of our interest and we calculate them by means of the general expression for the correlation function in the Lehmann's spectral representation:
\begin{eqnarray}
C_{m} (\omega) = \frac{1}{Z} \sum_{n n'} \frac{X^{m}_{nn'} X^{m}_{n'n}}{\omega + E_{n} - E_{n'}} [e^{- \beta E_{n}} + \xi e^{- \beta E_{n'}}],
\end{eqnarray}
where $X^{m}_{nn'}$ is a matrix element of the fermion ($\xi$ = 1) or spin ($\xi$=-1) operator and $Z$ is the partition function.
To avoid the extrapolation from Matsubara frequencies to real frequencies we use energy $\omega + i \delta$, where $\delta = \frac{\pi}{\beta} $. At low temperatures the small number of excited states contribute to the Boltzmann factor in Eq.(3). In our investigation we calculated these correlation functions for 40 excited states, which is enough to work in the temperature range 0 - 50 K with the Boltzmann factor smaller than 10$^{-6}$.

\begin{figure}[]
\includegraphics[width=0.34\textwidth,angle=0]{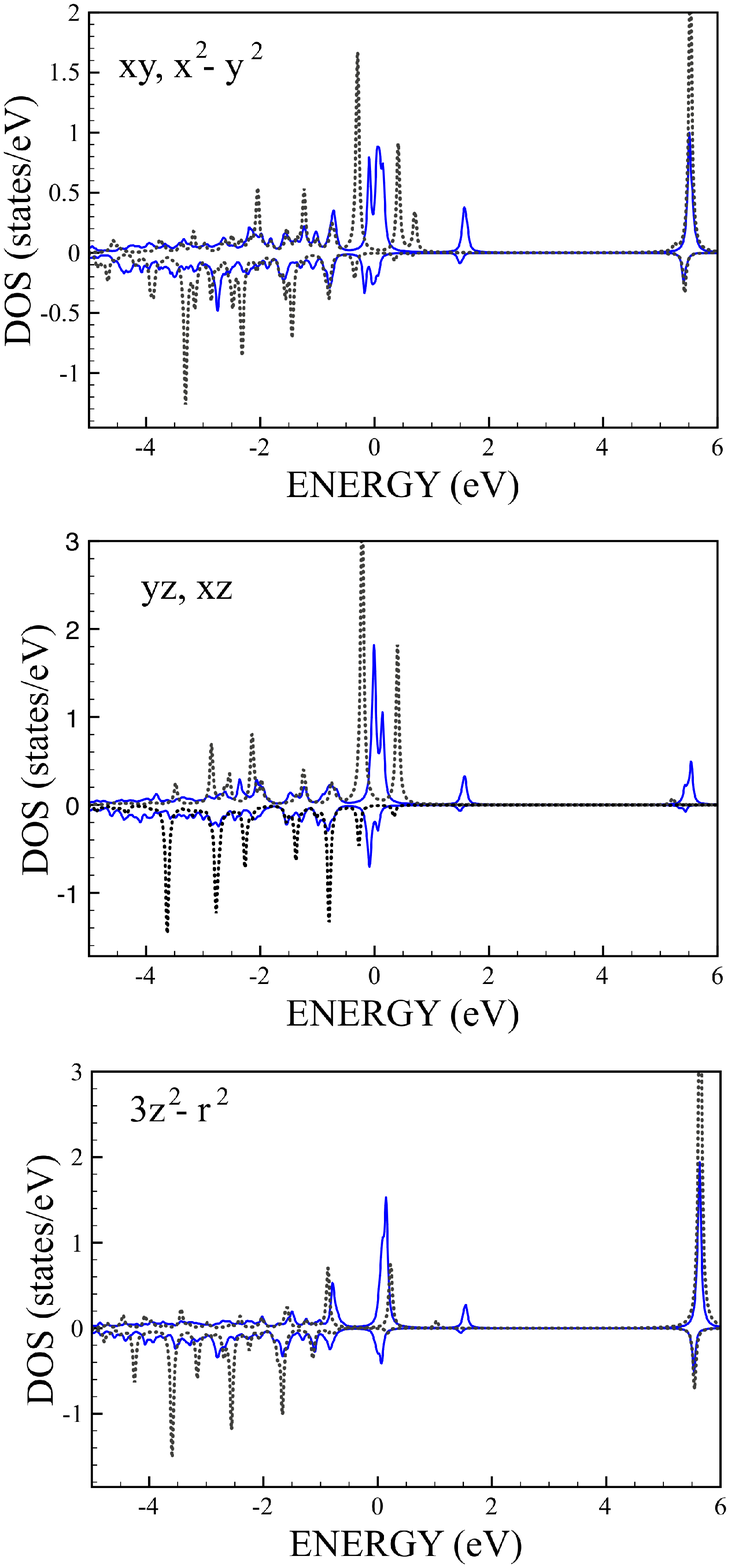}
\caption{(color online). Densities of states obtained from model calculations for U = 6.75 eV and J = 0.9 eV. The dashed gray and solid blue lines correspond to calculations with density-density Coulomb interaction and full U-matrix. The Fermi level is zero.}
\label{structures}
\end{figure}

In order to calculate the impurity Green's function (single-particle correlation function) we used $X^{m}_{nn'} = d_{nn'}$ in Eq.(3). These results for different types of Coulomb interaction matrices are presented in Fig. 3. The calculations with density-density and full Coulomb interaction matrices give slightly different values of the magnetic moment of 2.2 $\mu_{B}$ and 2.0 $\mu_{B}$, respectively. It means that the account of the non-diagonal elements of the U-matrix results in an additional suppression of the magnetic state. One can see that in the case of the density-density Coulomb matrix  the spectrum near the Fermi level is strongly renormalized due the spectral density transfer to the empty states at $\epsilon_{d}$ + U. In turn the account of non-diagonal elements of the U matrix results in new excitations of the system near the Fermi level.

The density of states obtained from the Anderson model calculations with the full U-matrix is in good agreement with the quasiparticle region of the differential conductance $dI/dV$ (Fig.1). There is a peak at about -0.05 eV below the Fermi level. This peak results from $xy$, $yz$, $xz$ and $x^2-y^2$ states which are strongly hybridized with conductance states of the Pt(111) substrate.  The width of the peak is 100 meV, which is much larger than that one obtained for a Kondo system. The states just above the Fermi level are mainly of $3z^2-r^2$ symmetry that is hybridized much less than other. Such a symmetry difference of the states near the Fermi level can explain the  different behaviour of the conductance at the external magnetic field \cite{Meier}. When the magnetic field is reversed the peak below the Fermi level strongly changes the intensity. It is due to the fact that the peak is originated from the partially occupied cobalt states that give the largest contribution to the total magnetic moment.

\begin{figure}[t]
\includegraphics[width=0.46\textwidth,angle=0]{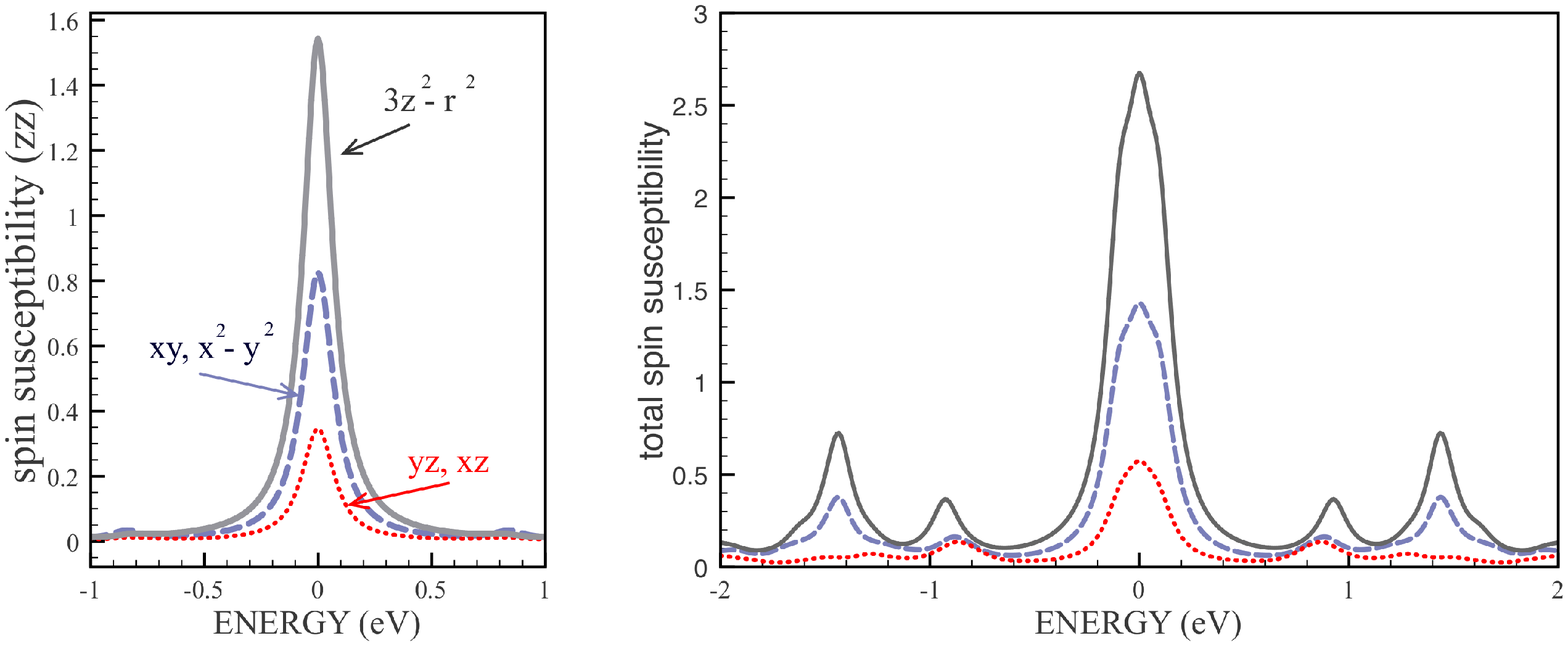}
\caption{(color online). Impurity z-z (left panel) and total spin-spin (right panel) correlation functions obtained from Anderson model calculations with full Coulomb interaction matrix. Solid gray, dashed light blue and dotted red lines correspond to spin susceptibilities of $3z^{2}-r^{2}$, $xy$ ($x^{2}-y^{2}$) and $yz$ ($xz$) orbitals, respectively.}
\label{structures}
\end{figure}

Another important source of information about an impurity system is the spin susceptibility that can be calculated by using Eq.(3) with $X^{m}_{nn'} = S^{z}_{nn'}$ and $\xi$=-1. These results are presented in Fig.4 (left panel).
One can see that all the susceptibilities yield peak which is mostly pronounced for $xy$($x^2-y^2$) and $3z^{2} - r^2$ states. It means that there is the local moment for these states\cite{Katanin}. As for the $yz$($xz$) state the susceptibility is nearly flat and we believe that the spin correlation function demonstrates an itinerant behaviour due to a strong hybridization between Co and Pt states. We have also calculated the correlation function for the linear combination $X^{m}_{nn'}X^{m}_{n'n} =S^{z}_{nn'}S^{z}_{n'n} + \frac{1}{2} (S^{+}_{nn'}S^{-}_{n'n} + S^{-}_{nn'}S^{+}_{n'n})$.
The resulting correlation functions presented in Fig.4 (right panel) demonstrate high-energy excitations which are mainly of $3z^2-r^2$ symmetry.
One can use the calculated spin susceptibilities in order to simulate inelastic excitations which are related with the transition matrix elements for the spin operators  \cite{Appelbaum, fernandez,Otte}. We left such an analysis for a future study.

In conclusion, the investigation of the electronic and magnetic properties of Co/Pt(111) system was carried out in the framework of the impurity Anderson model by using finite temperature Lanczos-solver. Using the obtained results we proposed an interpretation of the experimentally observed energy spectrum of the magnetic nanosystem on the surface in terms of strong many-body correlation effects.

{\it Acknowlegment.}
The hospitality of the Institute of Theoretical Physics of Hamburg University is gratefully acknowledged.
This work is supported by DFG Grant  No. SFB 668-A3 and the Cluster of Excellence "Nanospintronics" (Germany), 
RFFI 09-02-00431, RFFI 10-02-00546, RFFI 10-02-00046, the grant program of President of Russian Federation
 MK-1162.2009.2, the scientific program ``Development of scientific potential of universities'' N 2.1.1/779 and by the scientific program of the Russian Federal Agency of Science and Innovation N 02.740.11.0217 and N 2009-1.1-121-051-007.

\end{document}